\documentclass[]{emulateapj}
\submitted{Accepted for Publication in AJ}

\newcommand{\Ha }{H$\alpha$}
\newcommand{\HI}{H{\sc i}}
\newcommand{\kms} {km s$^{-1}$}
\newcommand{\siis}{[\mbox{S\,{\sc ii}}]$\lambda$6717}
\newcommand{\siio}{[\mbox{S\,{\sc ii}}]$\lambda$6731}
\newcommand{\nii}{[\mbox{N\,{\sc ii}}]$\lambda$6584}
\newcommand{\niib}{[\mbox{N\,{\sc ii}}]$\lambda$6548}
\newcommand{\NII}{[\mbox{N\,{\sc ii}}]}
\newcommand{\SII}{[\mbox{S\,{\sc ii}}]}
\newcommand{\OI}{[\mbox{O\,{\sc i}}]}
\newcommand{\OIII}{[\mbox{O\,{\sc iii}}]}

\shorttitle{Evidence for a Bar in NGC~2683}
\shortauthors{Kuzio de Naray, Zagursky, \& McGaugh}

\begin{document}
\title{Kinematic and Photometric Evidence for a Bar in NGC~2683}
\author{Rachel Kuzio de Naray\altaffilmark{1}}
\affil{Center for Cosmology, Department of Physics and Astronomy, University of California, 
  Irvine, CA 92697-4575}
\altaffiltext{1}{NSF Astronomy and Astrophysics Postdoctoral Fellow}
\email{kuzio@uci.edu}
\and
\author{Matthew J. Zagursky and Stacy S. McGaugh}
\affil{Department of Astronomy, University of Maryland, College Park,
  MD 20742-2421}
\email{mzagursk@umd.edu}
\email{ssm@astro.umd.edu}

\begin{abstract}
We present optical long-slit and SparsePak Integral Field Unit
emission line spectroscopy along with optical broadband and near IR
images of the edge-on spiral galaxy NGC~2683.  We find a multi-valued,
figure-of-eight velocity structure in the inner 45$\arcsec$ of the
long-slit spectrum and twisted isovelocity contours in the velocity
field.  We also find, regardless of wavelength, that the galaxy
isophotes are boxy. We argue that taken together, these kinematic and
photometric features are evidence for the presence of a bar in
NGC~2683.  We use our data to constrain the orientation and strength
of the bar.
\end{abstract}


\section{Introduction}
A large fraction of edge-on spiral galaxies have been observed to have
boxy or peanut-shaped (hereafter B/PS) bulges.  As the observations
have improved and the samples have grown, the percentage of galaxies
with these bulge shapes has increased from less than 20\%
\citep{Jarvis86,Shaw87,deSouza87} to up to 45\% \citep{Lutticke00a}.
Because the observed frequency of B/PS bulges is so high, significant
effort has been invested in determining the formation mechanism
responsible for these bulge shapes.

\citet{Binney85} and \citet{Rowley88} demonstrated that the accretion
of satellite galaxies could result in a galaxy bulge looking boxy or
peanut-shaped.  The currently favored formation mechanism, however, is
the buckling and subsequent vertical thickening of a bar
\citep[e.g.][]{Combes81,Combes90,Raha91}.

This is a plausible explanation, as bars are known to be common
features in disk galaxies.  \citet{Whyte02} found that 79\% of disk
galaxies are barred in the near-infrared and 74\% are barred in the
optical \citep[see also][]{Eskridge00,Marinova07}.  Barred galaxies
are also known to occur in a variety of environments. Using a sample
of 930 galaxies, \citet{vandenbergh02} showed that the presence
of a bar does not depend on whether the galaxy is in the field or is a
member of a group or a cluster.  Similarly, \citet{Verley07} studied the
bar frequency in isolated galaxies and found roughly equal numbers of
barred and unbarred systems.

To confirm the bar buckling hypothesis, bars must be detected in
edge-on galaxies with B/PS bulges.  While bars can be easily seen in
face-on galaxies, it is much more difficult, however, to unambiguously
photometrically confirm the presence of a bar in an edge-on galaxy.
This problem was simplified when \citet{Kuijken95} and
\citet{Merrifield99} demonstrated that bars in edge-on galaxies could
be detected kinematically from features like parallelograms and
figures-of-eight in the position-velocity diagrams.  Subsequent
observations and simulations \citep[e.g.][]{BF99,BA99,AB99,CB04,BA05}
have confirmed the relationship between B/PS bulges, complex
kinematics, and bars.

We have obtained the optical long-slit spectrum and SparsePak Integral
Field Unit (IFU) velocity field, as well as optical and near-IR images, 
of NGC~2683, an isolated, nearly edge-on Sb galaxy with a B/PS bulge.
NGC~2683 is highly inclined; \citet{Funes02} report an inclination of
78$\degr$, while others suggest a \textit{minimum} inclination of
80$\degr$ \citep[e.g.][]{Barbon75, Broeils94}.  The complex kinematic
and photometric features of the data are suggestive of a bar in the
galaxy.  Although previous authors have observed NGC~2683 and linked
its complex velocity structure and bulge shape with a bar, there has
been little discussion about the bar orientation.

In this paper we present our spectroscopic and photometric
observations as evidence that NGC~2683 contains a bar that is 6$\degr$
on the sky away from the galaxy major axis.  In $\S$ 2 we review the
data for NGC~2683 in the literature.  Our new observations and data
reduction are discussed in $\S$ 3.  We present the spectroscopic data
in $\S$ 4 and the photometric data in $\S$ 5.  We describe in each
section how these data support the hypothesis that a bar is present in
the galaxy, and use the features of the kinematic and photometric
signatures to constrain the orientation of the bar.  The results for
NGC~2683 are compared in $\S$ 6 with the results for M31 by
\citet{AthanassoulaBeaton}.  A comparison to M31 is insightful, as it
has photometric and kinematic characteristics, as well as an
inclination, that are similar to NGC~2683. Finally, a summary is
presented in $\S$ 7.

\section{Previous Observations}
Because of its prominent boxy/peanut-shaped bulge and interesting
kinematics, NGC~2683 has been a well-studied object.  In 1974, de
Vaucouleurs classified the bulge as peanut-shaped. \citet{Ford71}
obtained a long-slit spectrum of the galaxy and noted that outside of
the nucleus, the \nii\ emission was fainter than the \Ha\ emission,
whereas inside the nucleus, it was brighter than the \Ha\ emission.
\citet*{Barbon75} presented \textit{B-}band photometry and optical
emission line spectroscopy.  Based on the galaxy image, they found the
bulge to be peanut-shaped and the dust-obscured NW side of the galaxy
to be the near side. The spectroscopic data showed trailing spiral
arms, a \NII/\Ha\ ratio in the outer parts of the galaxy consistent
with what is seen in H{\sc ii} regions (\NII\ emission that is fainter
than \Ha\ emission), and strong evidence for deviations from circular
motions.  \citet{deSouza87} classified the bulge+disk shapes of 72
galaxies and categorized NGC~2683 as a BS-II galaxy: a system in which
the disk and bulge components are not sharply distinct and the bulge
is an elongated ellipsoid that is sometimes rectangular.  An \HI\
rotation curve of the galaxy was presented by \citet{Casertano91}, and
noncircular motions were noted.  \textit{J-} and \textit{K-}band
photometry, as well as \textit{(J-K)} color, eccentricity, position
angle and $\cos$~4$\theta$ profiles as functions of radius were
presented by \citet{Shaw93}, as were models of the observed luminosity
distribution.  \citet{Rubin93} found the \Ha\ and \NII\ emission to
show multi-valued, figure-of-eight velocity structure in the inner
60$\arcsec$ of the galaxy.  Light from bulge stars or foreground gas
warping out of the galaxy plane and into the line-of-sight were
suggested as possible explanations for the observed complex velocity
structure.  An \HI\ position-velocity map of NGC~2683 was presented by
\citet*{Broeils94}.

The link between boxy bulges and bars was investigated by
\citet{Merrifield99} using a sample of 10 galaxies, including
NGC~2683.  They found that the emission line kinematics of galaxies
with boxy bulges show complex velocity structure; the
position-velocity diagrams can be described as being X-shaped or
containing features similar to figures-of-eight or parallelograms.
The complex kinematics of NGC~2683 were also observed by
\citet{Pompei99} who noted the presence of a counter-rotating stellar
system and two gas components, one of which was rotating in the same
sense as the stellar component, and the other rotating in the opposite
direction.  A nuclear bar close to the minor axis or a merger event
were two scenarios given to explain the counter-rotation.
Additionally, \citet{Pompei99} observed the bulge isophotes to be boxy
inside 50$\arcsec$.  Optical and NIR photometry also led
\citet{Lutticke00a,Lutticke00b} to classify the bulge of NGC~2683 as
box-shaped.  Stellar and ionized gas velocity curves as well as
velocity dispersions were presented by \citet{VegaBeltran01}.  They,
too, detected kinematically distinct stellar and gas components, but
remarked that they were unable to resolve the fast and slow rotating
components in the gas.  \citet{Funes02} presented a position-velocity
diagram for NGC~2683 derived from \Ha, \nii, and [\mbox{O\,{\sc
iii}}]$\lambda$5007 emission.  Due to the low \textit{S/N} of their
data, the figure-of-eight shape in the position-velocity diagram was
faint, but they did observe two spatially distinct gas components.

\textbf{ }

\section{New Observations and Data Reduction}
\subsection{Spectroscopy}
We observed NGC~2683 with the RC Spectrograph on the Kitt Peak
National Observatory (KPNO) 4-meter telescope during the nights of
2007 October 16 and 18. We used the T2KB CCD with the 860 line
mm$^{-1}$ grating in second order, centered near \Ha.  The slit width
was 1.5$\arcsec$, giving $\sim$ 1.0 \AA\ spectral resolution and a
spatial scale of 0.69$\arcsec$ pixel$^{-1}$.  We centered the slit on
a nearby bright star and then offset to the optical center of the
galaxy.  Two 600 s exposures were taken with the slit aligned along
the major axis of the galaxy (position angle $=$ 41.5$\degr$).  A
HeNeAr lamp was observed before and after each science exposure to
provide wavelength calibration.

We also observed NGC~2683 with the SparsePak IFU on the 3.5-meter WIYN
$\footnote{Based on observations obtained at the WIYN Observatory.
The WIYN Observatory is a joint facility of the University of
Wisconsin-Madison, Indiana University, Yale University, and the
National Optical Astronomy Observatory.}$ telescope at KPNO on the
night of 18 February 2009.  SparsePak is a 70$\arcsec$ $\times$
70$\arcsec$ fixed array of 82 5$\arcsec$ diameter fibers.  We used the
STA1 CCD with the 316@63.4 grating in eighth order, centered near
H$\alpha$, giving a 40 km s$^{-1}$ velocity resolution.  The SparsePak
array was aligned with the major axis of the galaxy and five pointings
were used to cover the entire length of the galaxy.  Each exposure was
1200 s, and two exposures were taken at each pointing.  A ThAr lamp
was observed to provide wavelength calibration.

The spectral data were reduced using standard reduction routines in
IRAF$\footnote{IRAF is distributed by the National Optical Astronomy
Observatory, which is operated by the Association of Universities for
Research in Astronomy (AURA), Inc., under agreement with the National
Science Foundation.}$.  The data were bias-subtracted and flattened.
The IRAF task \texttt{dohydra} was used to extract the IFU spectra.
The galaxy spectra were wavelength-calibrated using a wavelength
solution created from the observations of the comparison HeNeAr and
ThAr lamps.  To increase the signal-to-noise and remove cosmic rays,
the two galaxy frames per pointing were combined. In general, the
galactic emission lines in the spectra are much stronger than the
night-sky emission lines, and the night-sky lines bracket, rather than
overlap, the galaxy lines.  Sky subtraction was not performed for
either data set and we subsequently used the night-sky emission lines
as the reference wavelengths \citep{Osterbrock96} by which the
velocities of the galactic emission lines were measured.  The
velocities were measured by fitting Gaussians to both the sky lines
and the five galactic emission lines of interest: \Ha, \niib, \nii,
\siis, \siio.  There was less scatter between the measured galactic
emission line velocities when using the night-sky calibration than
either the HeNeAr or ThAr calibrations.

\begin{figure*}[ht]
\epsscale{0.8}
\plotone{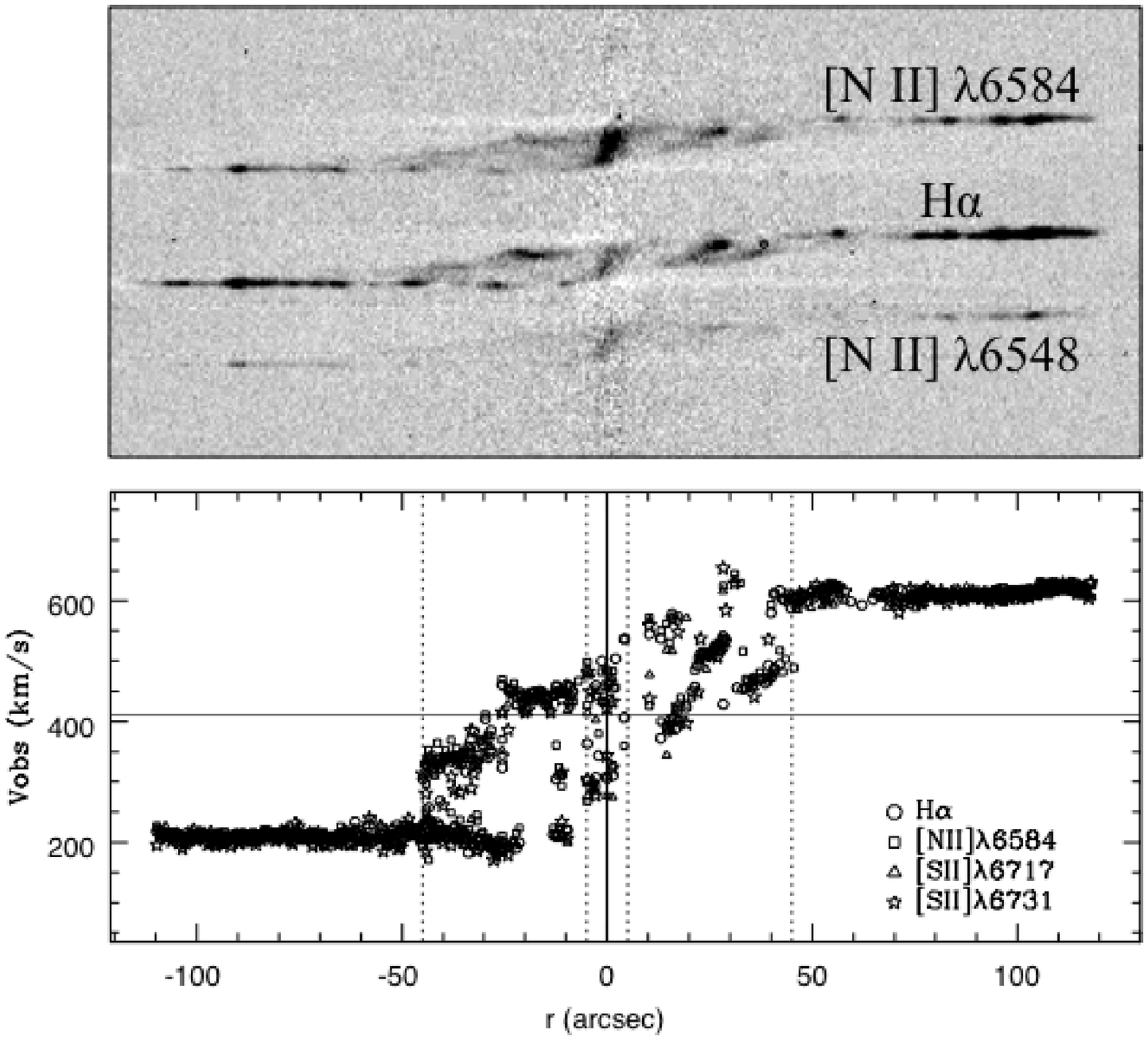}
\caption{\textit{Top:} Observed \Ha\ and \NII\ region of the long-slit
position-velocity diagram.  The emission traces two parallelogram
shapes: an outer parallelogram extending to about $\pm$45$\arcsec$,
and a compact, inner parallelogram extending to about
$\pm$5$\arcsec$. \textit{Bottom:} Observed velocities of the \Ha,
\nii, \siis, and \siio\ emission lines.  The error on the velocity at
a given position is equal to the scatter in the measured velocities of
the observed emission lines, $\sim$10 \kms.  At radii $\lesssim$
45$\arcsec$, the velocities are multi-valued.  The measured velocities
occupy the two forbidden quadrants of the PVD; there is red-shifted
emission on the blue-shifted side of the galaxy (upper left quadrant)
and blue-shifted emission on the red-shifted side of the galaxy (lower
right quadrant).}
\end{figure*}

\subsection{Photometry}
The galaxy was also imaged at the KPNO 2.1-meter telescope during the
night of 2007 March 17.  The galaxy was imaged in $B$, $V$, $R$, and
$I$, with total exposure times of 600 s, 600 s, 300 s, and 300 s,
respectively.  Two exposures were taken in each band to correct for
cosmic rays.  The T2KB CCD was used, and the spatial resolution was
0.305$\arcsec$ pixel$^{-1}$.  These data were taken under
non-photometric conditions.  The optical broad-band images were
bias-subtracted, flattened, and combined using standard IRAF
routines. 2MASS \textit{J-} and \textit{K-}band images of NGC~2683 are
also available, and were downloaded from the on-line 2MASS catalog
\citep{2MASS}.

\section{Kinematic Signatures}
\subsection{Long-slit Position Velocity Diagram}
In Figure 1 we show the observed \Ha\ and [\mbox{N\,{\sc
ii}}]$\lambda$$\lambda$6548, 6584 region of the long-slit
position-velocity diagram (PVD) of NGC~2683.  The PVD shows complex
structure out to radii of $\sim$ 45$\arcsec$, almost half of its
entire length.  The \Ha\ and \NII\ emission (as well as the \siis\ and
\siio\ emission that are not shown) trace a parallelogram shape
extending to $\sim$ 45$\arcsec$ on each side. The edges of this region
are relatively bright compared to the faint interior emission.  At
radii $\lesssim$ 45$\arcsec$, the velocities are double, sometimes
triple, valued.  The maximum velocity reached is comparable to the
velocities reached in the outer parts of the galaxy.  The
parallelogram velocities are seen to populate the ``forbidden''
quadrants of the PVD; there are red-shifted velocities on the
blue-shifted side of the galaxy and vice versa.

There is also a second parallelogram feature seen in the PVD.  In
approximately the inner $\pm$ 5$\arcsec$ there is a steeper and more
compact parallelogram that is brighter in the \nii\ emission than \Ha.
The velocities reached in this region do not exceed the velocities in
the more extended parallelogram or those of the outer disk, and there
is a smooth transition between the compact, inner and extended, outer
parallelogram-shaped features.

This complex ``figure-of-eight'' PVD can be explained by the presence
of a bar in the galaxy.  In a series of papers, \citet[][hereafter
BA99]{BA99}, \citet[][hereafter AB99]{AB99}, and \citet[][hereafter
BA05]{BA05} used simulations of edge-on ($i$ = 90$\degr$) galaxies to
demonstrate that features in PVDs like those shown in Figure 1 can be
explained by gas (and stars) moving in the \textit{x$_{1}$} and
\textit{x$_{2}$} families of periodic orbits of barred potentials.
Each of these orbit families produces unique signatures in the PVD
and, by virtue of the orientations of the orbits with respect to the
bar, the appearance of the PVD features can be used to put constraints
on the position angle of the bar.

\defcitealias{BA99}{BA99}
\defcitealias{AB99}{AB99}
\defcitealias{BA05}{BA05}

\begin{figure*}[ht]
\plotone{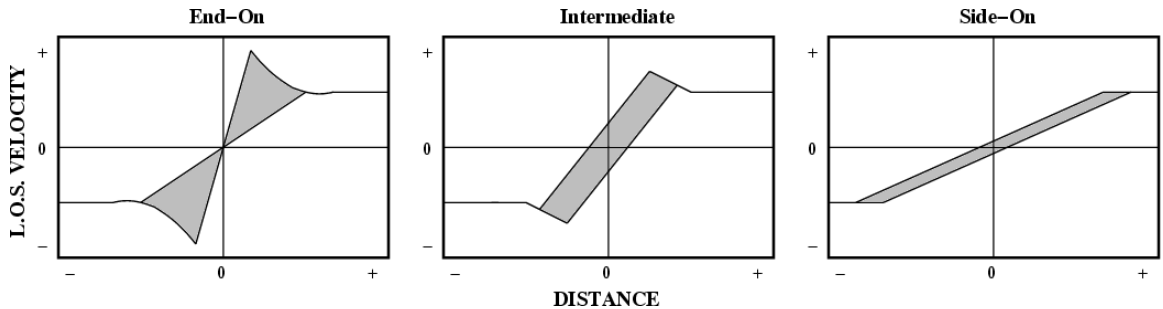}
\caption{Schematic representation of the shape of the
position-velocity diagram produced by \textit{x$_{1}$} orbits for
different lines of sight.  The major axes of both the bar and
\textit{x$_{1}$} orbits are parallel to the observer's line-of-sight
when the bar is viewed end-on, and are perpendicular to the observer's
line-of-sight when the bar is viewed side-on.}
\end{figure*}

\citetalias{BA99}, \citetalias{AB99}, and \citetalias{BA05} showed
that in edge-on galaxies \textit{x$_{1}$} orbits produce extended
features in the PVD that range from bow tie-shaped to
parallelogram-shaped depending on the orientation of the bar with
respect to the observer's line-of-sight (see Figure 2).  The
\textit{x$_{1}$} orbits are elongated parallel to the major axis of
the bar, so that when the bar is viewed end-on, the line-of-sight is
parallel to the major axis of both the bar and \textit{x$_{1}$}
orbits.  Viewed at this angle, the ``fast'' parts of the elongated
\textit{x$_{1}$} orbits are pointing directly toward or away from the
observer and very high radial velocities at small projected distances
are observed.  This produces a bow tie-shaped feature in the PVD.

When the bar is viewed side-on, the elongated axis of the
\textit{x$_{1}$} orbits is perpendicular to the line-of-sight.  The
``slow'' parts of the \textit{x$_{1}$} orbits are thus pointing
directly toward or away from the observer, and as a result, much lower
radial velocities are observed.  In addition, because the long axis of
the orbits is across the line-of-sight, these observed velocities will
extend to larger projected distances than when the bar (and
\textit{x$_{1}$} orbits) is viewed end-on.  The resulting feature in
the PVD is therefore a relatively thin, diagonal band of velocities
extending to large projected distances.

When the bar in an edge-on galaxy is not viewed directly end-on or
side-on, the PVD will contain a parallelogram-shaped feature with
velocities appearing in the ``forbidden'' quadrants.  This velocity
structure arises because the orbits are elongated and not circular.
The line-of-sight does not fall along either the major axis or minor
axis of the orbits. Thus, if they are viewed at an angle other than
end-on or side-on, the positions along the orbits where the tangential
velocity is greatest (and radial velocity is zero) will be offset from
the center.  This means that the velocities transition from being
blue-shifted to red-shifted, and vice versa, at radii greater than
zero.  This leads to red(blue)-shifted velocities being observed on
the blue(red)-shifted side of the galaxy, thus falling in the
``forbidden'' quadrants.  The closer the bar is to end-on, the sharper
and more peaked the parallelogram is at small projected radii (and the
more it looks like a bow tie).  The interior of these PVDs also tend
to be brighter than the edges.  As the bar angles closer to side-on,
the corners of the parallelogram round out and the parallelogram
becomes thinner as it extends to larger projected distances and rises
to lower radial velocities.

In the observed PVD of NGC~2683, the broad ($\pm$45$\arcsec$)
parallelogram shape corresponds to the \textit{x$_{1}$} feature.  The
simulated PVDs of \citetalias{BA99}, \citetalias{AB99}, and
\citetalias{BA05} can serve only as a \textit{guide} to the bar
orientation in our galaxy because (a) NGC~2683 is not exactly edge-on
($i$ $\gtrsim$ 78$\degr$), and (b) the models are not tailored to
specifically match NGC~2683 (bar strength, pattern speed, mass
distribution, etc).  With that caveat in mind, both the bright
emission along the edges of the \textit{x$_{1}$} parallelogram and a
lack of clear central peaks suggest that the bar is oriented at an
intermediate angle (perhaps between 10$\degr$ and 45$\degr$) on the
sky away from the galaxy major axis and therefore viewed closer to
side-on than end-on.

At the center of the observed PVD of NGC~2683 is the signature of the
\textit{x$_{2}$} orbits: a compact feature that is brighter in \NII\
emission than \Ha\ emission \citep[Figure 1; see also][]{Ford71}.
This relatively large \NII\ emission may be caused by the shocks that
develop at the transition between the \textit{x$_{1}$} and
\textit{x$_{2}$} orbits (Baldwin, Phillips, \& Terlevich 1981; Dopita
\& Sutherland 1996; AB99), as well as the AGN that is known to be at
the center of NGC~2683 \citep{Irwin}.  Without additional diagnostic
emission lines such as \OI\ or \OIII, we cannot say for certain that
it is the AGN rather than shocks that is causing the bright \NII\
emission.  It is not uncommon for barred galaxies to host
AGN. \citet{Philips83} observed a number of relatively face-on SB
galaxies with large \NII\//\Ha\ ratios and confirmed the presence of
active nuclei.  \citet{Knapen00} have also found that galaxies with
active centers are more often barred (79\%) than galaxies without
active nuclei (59\%) \citep[see also][]{Laurikainen04}.

The \textit{x$_{2}$} orbits are elongated perpendicular to the bar
(making them also perpendicular to the \textit{x$_{1}$} orbits) and
are less radially extended than the \textit{x$_{1}$} orbits.  Because
the major axes of the \textit{x$_{2}$} and \textit{x$_{1}$} orbits are
perpendicular, when one orbit family reaches its maximum radial
velocity, the other is at its minimum.  Thus, the radial velocities of
the \textit{x$_{2}$} orbits peak when the bar is viewed side-on and
are lowest when the bar is viewed end-on.

Rather than using the shape of the \textit{x$_{2}$} feature in the
PVD, \citetalias{AB99} advocate the ratio of the maximum observed
radial velocity of the \textit{x$_{2}$} orbits to the maximum velocity
observed in the outer parts of the galaxy ($V_{x2}$/$V_{disk}$) as the
best measure of bar viewing angle.  When the bar is viewed close to
end-on, the maximum velocity of the \textit{x$_{2}$} orbits will be
low compared to the outer disk (the line-of-sight is perpendicular to
the elongated axis of the \textit{x$_{2}$} orbits and the ``slow''
parts of the orbit are observed); $V_{x2}$/$V_{disk}$ $\lesssim$ 1 in
this case.  The ``fast'' parts of the \textit{x$_{2}$} orbits are
observed when the bar is viewed side-on; it therefore follows that
$V_{x2}$/$V_{disk}$ $\gtrsim$ 1 in this orientation.  In NGC~2683, we
find that $V_{x2}$/$V_{disk}$ $\approx$ 0.6. This value of
$V_{x2}$/$V_{disk}$ suggests that the bar is viewed nearer to end-on
than side-on.

\begin{figure*}[ht]
\epsscale{0.8}
\plottwo{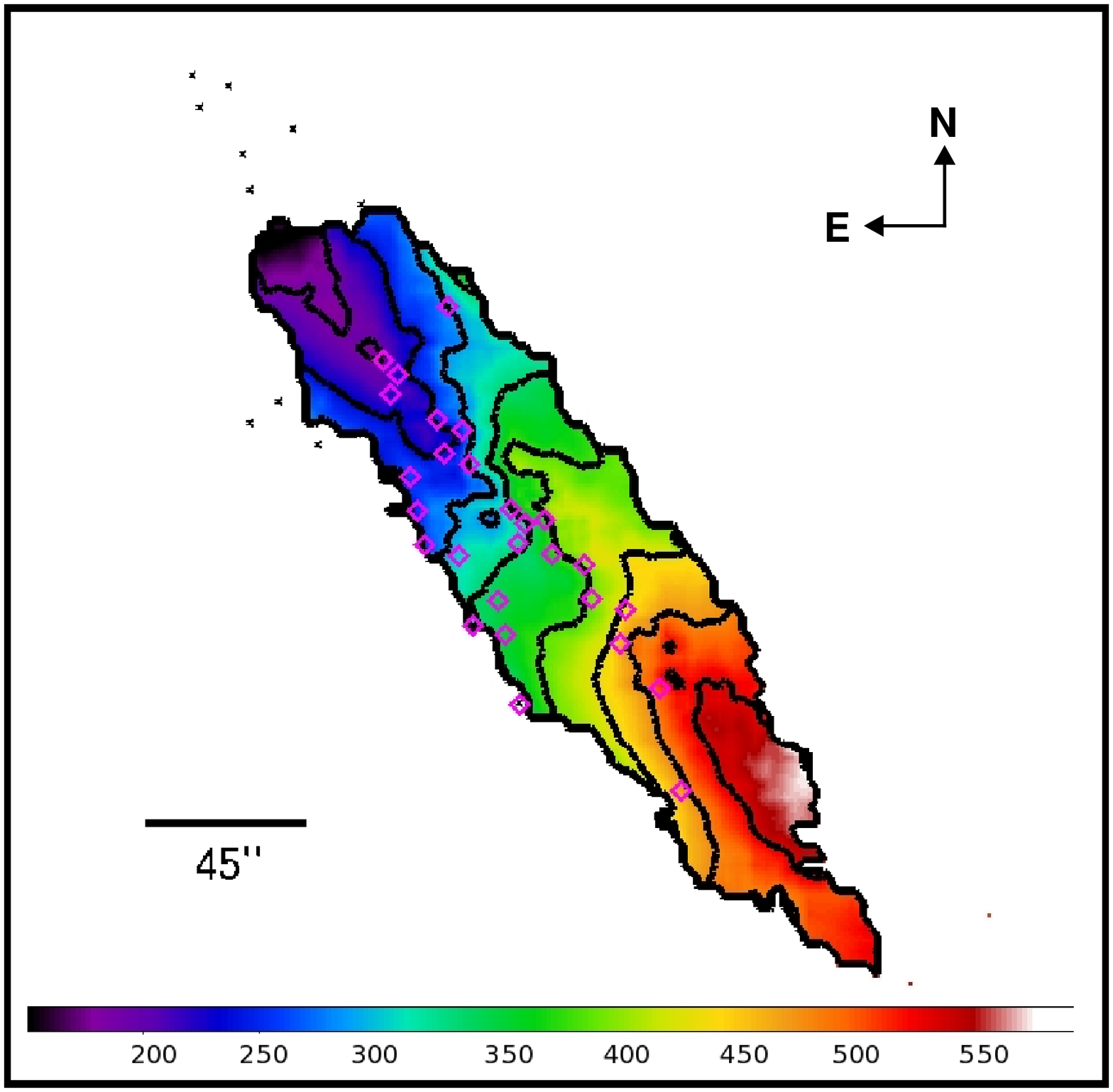}{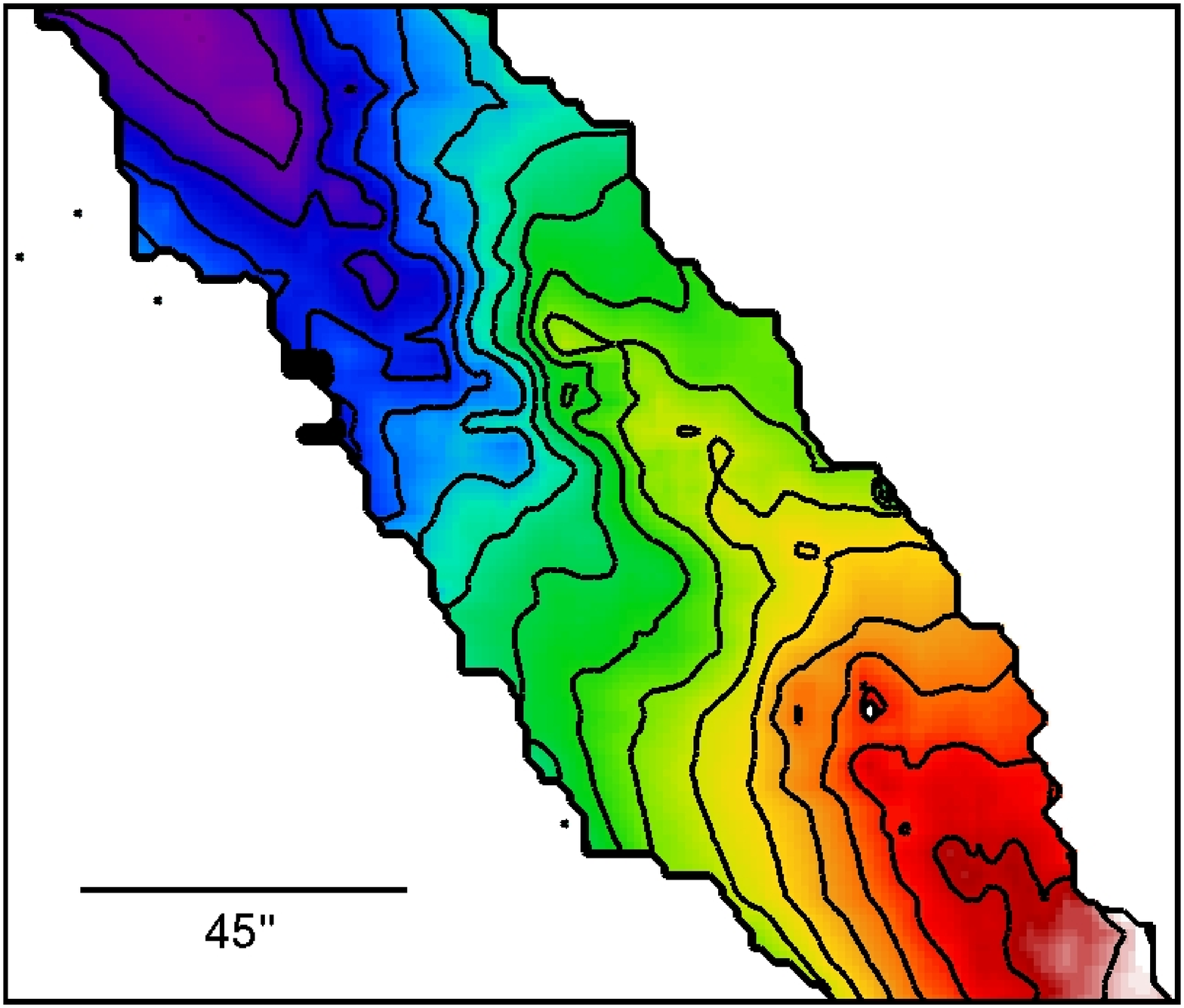}
\caption{Observed \Ha\ SparsePak velocity field of NGC~2683.  The
entire velocity field is shown in the left panel while a zoomed-in
version with the same color/velocity scale is shown on the right.  The
galaxy major and minor kinematic axes are clearly skewed from
perpendicular.  Diamonds in the left panel indicate fibers containing
multi-valued \Ha\ velocities.  In the full velocity field, isovelocity
contours are spaced in 50 \kms\ intervals. To highlight the twist of
the minor axis, the isovelocity contours are spaced in 25 \kms\
intervals in the right panel.  In both panels, a 45$\arcsec$ line
indicates the radius of the parallelogram in the long-slit PVD in
Figure 1.  North is up and East is to the left.  (A color version of
this figure is available in the online journal.)}
\end{figure*}

\begin{figure}
\epsscale{0.8}
\plotone{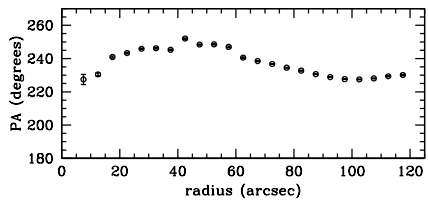}
\caption{Variation in position angle (PA) across the velocity field.}
\end{figure}

This result contrasts with the other indicators we have, both
photometric and kinematic, that suggest a more side-on orientation.
The $V_{x2}/V_{disk}$ indicator advocated by \citetalias{AB99} appears
to be one of the more promising indicators of bar orientation.
However, it must depend on the details of the disk mass distribution.
Some galaxies have a distinct peak in their rotation curves before
declining to a flat level while others rise continuously.
Consequently, $V_{disk}$ of this particular galaxy may not correspond
well to that of the AB99 model.  More generally, it may be an
indication that we still do not have a complete understanding of the
complex kinematics in NGC 2683.

\subsection{SparsePak IFU Velocity Field}
In Figure 3 we show the observed \Ha\ SparsePak velocity field of
NGC~2683.  The major and minor kinematic axes are not perpendicular
and there is an \textsf{S}-shaped twist to the minor axis.  These are
characteristic of velocity fields with oval distortions or bars
\citep[e.g.][]{Bosma81}.  Similar \textsf{S}-shaped velocity fields
can be seen in observations of more face-on barred galaxies such as
NGC~6300 \citep[\textit{i}=52$\degr$,][]{Buta87} and NGC~5383
\citep[\textit{i}=50$\degr$,][]{Peterson78}.  As in the long-slit
spectrum, double and triple-valued \Ha\ velocities (as well as \NII\
and \SII\ velocities) are detected in some of the IFU fibers.  The
multi-valued velocities are more striking in the long-slit spectrum
than in the velocity field.

Analysis of the velocity field with the tilted-ring fitting program
\texttt{ROTCUR} \citep{Teuben,Begeman} shows that the position angle
(PA) varies by $\sim$ 25$\degr$ across the velocity field, with the
largest deviation from the PA of the outer galaxy occuring at $\sim$
45$\arcsec$, the length of the \textit{x$_{1}$} parallelogram in the
long-slit PVD (see Figure 4).  If we assume that there should be a
steep velocity gradient (i.e. the isovelocity contours bunch up)
across the bar and that the central minor axis velocity contours run
along the length of the bar, then the center of the velocity field
qualitatively indicates (visually) that the bar is angled not far
(perhaps $\sim$ 15$\degr$) from the galaxy major axis on the sky,
running from northeast to southwest.  This is supported by the change
in PA determined by \texttt{ROTCUR}; the bar is a maximum of 25$\degr$
from the major axis.  This angle implies that the bar is viewed more
side-on than end-on, in agreement with the constraints from the
parallelogram in the long-slit PVD produced by \textit{x$_{1}$}
orbits.

\begin{figure}
\plotone{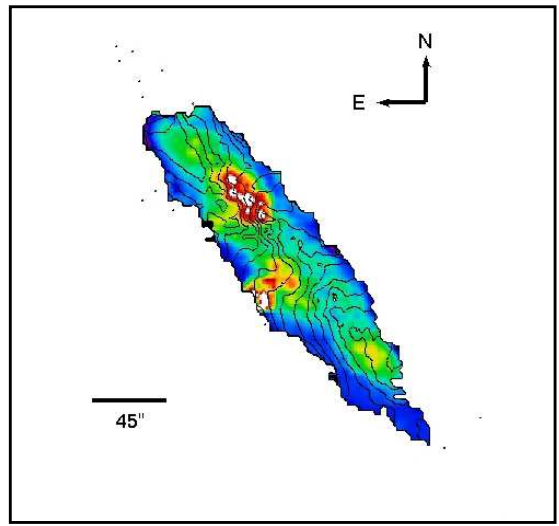}
\caption{\Ha\ image of NGC~2683 created from the SparsePak IFU data
with velocity contours overlaid.  The brightest and most concentrated
\Ha\ emission coincides with the ends of the bar. (A color version of
this figure is available in the online journal.)}
\end{figure}

In Figure 5 we show the \Ha\ image of NGC~2683 constructed from the
IFU data with isovelocity contours from the velocity field
overplotted.  We find that while there is diffuse emission throughout
the galaxy, the brightest concentrations of emission coincide with the
bends in the \textsf{S}-shape (the ends of the bar).  The NE (upper)
end of the bar is slightly brighter than the SW (lower) end and there
is no strong emission along the length of the bar.

\begin{figure*}[ht]
\plotone{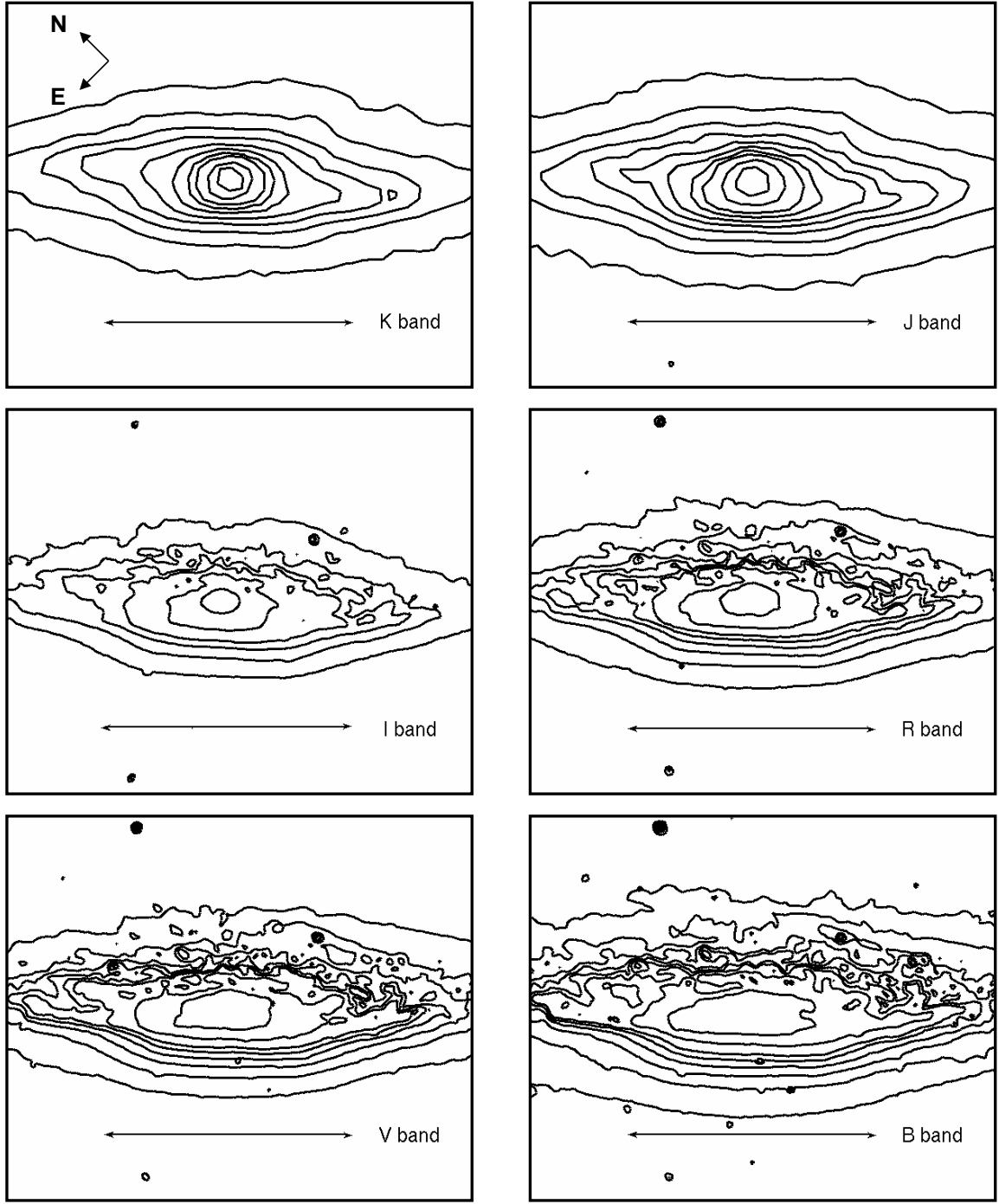}
\caption{Contour plots of the bulge region of NGC~2683, ordered by
wavelength from longest to shortest.  For comparison, isophotes are at
the same relative levels for each frame.  The bulge of NGC~2683 is
boxy: the isophotes are parallel to the major axis and do not show a
pinch along the minor axis.  The arrows indicate $\pm$45$\arcsec$, the
length of the \textit{x$_{1}$} extended parallelogram feature in the
position-velocity diagram. North is towards the upper left corner,
East to the lower left corner.  In the optical images there is
significant dust on the northwest side of the galaxy.}
\end{figure*}

\begin{figure*}[ht]
\plotone{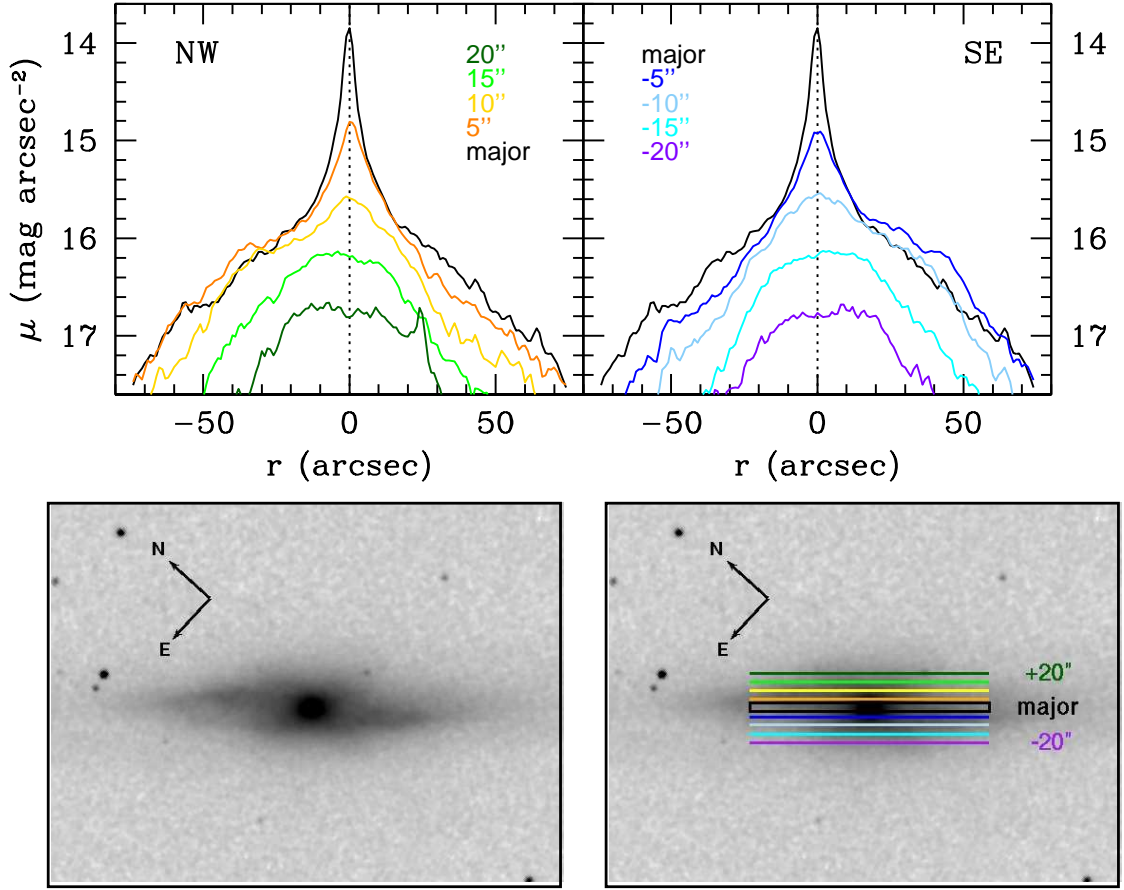}
\caption{$K$-band surface brightness measured along
5$\arcsec$$\times$140$\arcsec$ slits positioned parallel to the major
axis.  There is an enhancement in the light profile in the N$-$S
direction, suggesting the presence of a bar at a mild angle with
respect to the major axis.  (A color version of this figure is
available in the online journal.)}
\end{figure*}

\begin{figure*}[ht]
\plotone{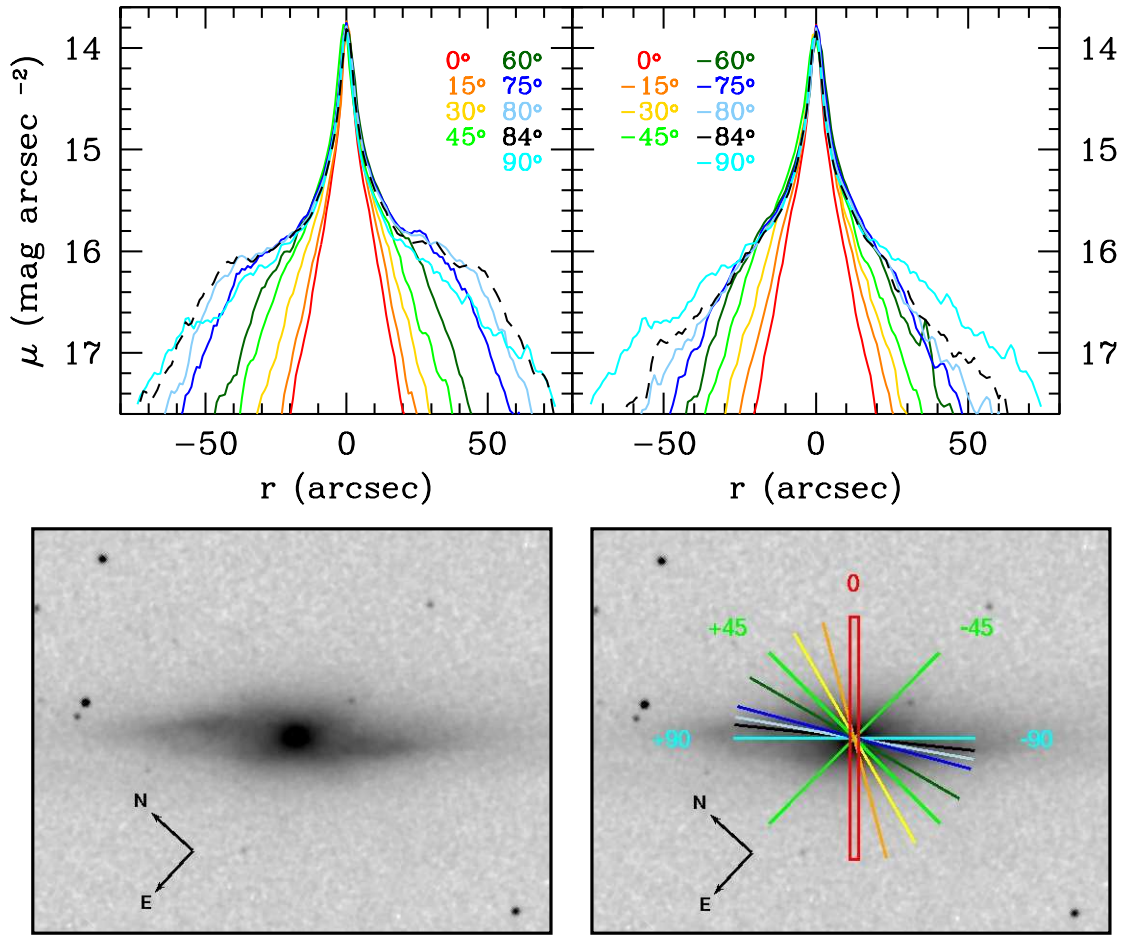}
\caption{$K$-band surface brightness measured along
5$\arcsec$$\times$140$\arcsec$ slits oriented at different position
angles.  There is an enhancement in the light above the level of the
major axis along the +84$\degr$ slit. (A color version of this figure
is available in the online journal.)}
\end{figure*}

Strong \Ha\ emission at the bar ends and relatively weak emission
along its length has also been seen in the relatively face-on galaxies
NGC~1530 \citep{Zurita04}, NGC~5383 \citep{Sheth00}, and NGC~6300 and
NGC~1433 \citep{Buta01}.  Similar to what is seen in NGC~2683,
\citet{Zurita04} found that the brightest \Ha\ emission in NGC~1530
coincides with the regions of lowest velocity gradient perpendicular
to the bar (regions were the isovelocity contours are not bunched up).

\section{Photometric Signatures}
Boxy/peanut-shaped bulges in edge-on galaxies have been shown to
correlate with the presence of a bar
\citep[e.g.][]{Combes81,Combes90,BF99}.  It is thought that these
bulge shapes are a result of a bar that has buckled and thickened due
to vertical instabilities \citep[e.g.][]{Combes90}.  Depending on the
bar strength and the orientation of the bar with respect to the
line-of-sight, the bulge may look round (the bar is viewed end-on),
boxy (the bar is viewed at an intermediate angle), or peanut-shaped
(the bar is viewed side-on) \citep[e.g.][]{CB04,BA05}.

In Figure 6, we show contour plots of the bulge region of the
\textit{B, V, R, I, J,} and \textit{K} images of NGC~2683.
\citet{Lutticke00a} define peanut-shaped bulges as having isophotes
that pinch inwards along the minor axis on both sides of the galaxy
major axis, and boxy bulges as having isophotes that remain parallel
to the major axis.  Applying these definitions to Figure 6, it is
clear that the bulge of NGC~2683 is boxy.  This is most easily seen in
the \textit{K-}band image of the galaxy where the obscuration from
dust is at a minimum.  On both sides of the major axis, the isophotes
are parallel and remain flat as they cross the minor axis.  Even as
the galaxy is viewed through progressively bluer filters and the
effects of dust on the northwest side of the galaxy become more
severe, the isophotes on the less-obscured side of the galaxy do not
show signs of pinching along the minor axis that are characteristic of
a peanut-shaped bulge.  The classification of the bulge shape is
independent of wavelength \citep[see also][]{Lutticke00a}.

Two of us (RKD and SSM) made independent ellipse fits with different
fitting routines (\texttt{ellipse} in IRAF and ARCHANGEL
\citep{Schombert07}).  While these ellipse fits are broadly
consistent, they give slightly different results for the
boxiness/pointiness parameter $b_4$.  Thorough examination of the
images and residuals of the fits shows that this can be traced to
isophotes that are both boxy and pointy.  This is most apparent in the
$J$ and $K$-band images in Figure 6.  Along the minor axis of the
bulge, the isophotes are flatter than a pure ellipse, indicating a
boxy bulge.  Along the major axis, the isophotes become pointy, so
that the entire isophote shape is both pointy and boxy.  We interpret
the pointiness to be due to the bar itself extending out into the
disk; see Figures 7 and 8.  The length of the boxy region,
corresponding to the thick inner part of the bar, is about
$\pm$23$\arcsec$ in the $K$-band.  The pointy isophotes extend to
roughly $\pm$54$\arcsec$ and correspond to the thin outer part of the
bar.

It is not uncommon for barred galaxies to have features that connect
to the ends of the bar such as ansae (``handles'') or rings
\citep[e.g.][]{Martinez07,Buta01}.  Because NGC~2683 is so highly
inclined, it is difficult to determine if these structures are also
present.

The boxy bulge shape of NGC~2683 indicates that the bar is not so
strong as to induce the pinched isophotes of a peanut and that the bar
is positioned at an intermediate viewing angle.  To determine the
general orientation of the bar, we plot in Figure 7 the observed
surface brightness measured along 5$\arcsec \times 140\arcsec$ slits
positioned parallel to the galaxy major axis.  We find that the
surface brightness profiles are asymmetric around the minor axis.
Specifically, in the slices above the major axis (the northwest
slices), there is a ``hump'' in the profile indicating excess light to
the left (north) of the galaxy minor axis and in the slices below the
major axis (the southeast slices), the ehancement in the light is to
the right (south) of the galaxy minor axis.  We interpret this
asymmetry in the light profiles as an indication that the bar is
aligned in a N-S direction, consistent with the orientation derived
from the SparsePak velocity field.  Consistent with the approximate
lengths of the boxy and pointy isophote regions, the excess light
begins around $\pm$ 20$\arcsec$$-$25$\arcsec$, peaks around
45$\arcsec$, and drops off between 50$\arcsec$ and 60$\arcsec$.  Based
on the photometry, we estimate the length of the bar to be $\sim$
55$\arcsec$.

In Figure 8, we confirm this N-S orientation by plotting surface
brightness profiles along slits oriented at different position angles.
We define 0$\degr$ as the minor axis and 90$\degr$ as the major axis.
There is an enhancement in the light along the slits rotated
counterclockwise (positive angles) from the galaxy minor axis.  The
humps in the profile reach a maximum above the level of the light
along the major axis at a position angle of +84$\degr$ from the minor
axis (or, equivalently, 6$\degr$ from the major axis).  This angle
coincides with the asymmetric, elongated isophotes seen most clearly
in the \textit{K}-band image.  In agreement with the kinematic data,
the humps in the surface brightness profiles extend to $\sim$
45$\arcsec$$-$50$\arcsec$.  Based on the photometric signatures, we
conclude that the bar is oriented $\sim$ 6$\degr$ from the galaxy
major axis on the sky, running from north to south.

\section{Comparison to M31}

It is instructive to compare the observations of NGC~2683 with
observations of other galaxies having similar inclinations ($i$
$\gtrsim$ 78$\degr$) and bar orientations ($\varphi_{phot}$ $\sim$
6$\degr$, $\varphi_{kin}$ $\sim$ 10$\degr$ $-$ 25$\degr$).  One such
example is M31.  \citet{AthanassoulaBeaton} recently used kinematic
and photometric data to argue for the presence of a bar in the galaxy
\citep[see also][]{Lindblad56,Stark94}.  They find that M31 ($i$ =
77$\degr$) contains a medium-strength bar that is on the order of
$\varphi$ $\sim$ 7$\degr$ on the sky away from the galaxy major axis
(PA$_{disk}$ = 38$\degr$, PA$_{bar}$ $\sim$ 45$\degr$).

The observed PVDs of NGC~2683 and M31 show similar complex kinematics.
The observed \HI\ PVD of M31 (their Figure 9) is more bow tie-shaped
than the long-slit \Ha\ PVD of NGC~2683.  As discussed in $\S$4.1, in
edge-on galaxies, the PVD appears more bow tie-like as the bar angles
farther away from the major axis.  Taken at face value, the observed
PVDs would therefore suggest that the bar in M31 is seen more end-on
than the bar in NGC~2683.  The simulations described in $\S$4.1,
however, are \textit{ideal} and in reality, the observed kinematics
are highly influenced by physical conditions in the galaxy, such as
the presence of dust or the density of the gas.  Without further
modeling, we cannot say whether the differences in the PVDs originate
from these physical differences or from different bar orientations or some
combination of both.  

The isophotes seen in the images of M31 (their Figure 2) are similar
to the isophotes of the $K$-band image of NGC~2683; the isophotes are
boxy rather than pinched in a peanut shape.  Based on isophote shapes
in $N$-body simulations tailored to the orientation of M31, they argue
that the absence of peanut-shaped isophotes means the bar is not a
strong bar \citep[see also][]{Lutticke00b,Athan02}.  By this standard, the
bar in NGC~2683 is not strong.  Similar to NGC~2683, M31 has
elongated isophotes that are angled away from the major axis and
indicate the PA of the bar.  Additionally, they find similar
asymmetric enhancements in the light profile in slits parallel to the
major axis (their Figures 5 and 6).

It is encouraging that there are so many similarities between the
observations of these two galaxies.  Bars in nearly edge-on galaxies
cannot be unambiguously visually confirmed as such, but strong
evidence for their presence comes from the comparison of observed
kinematic and photometric signatures to the results of numerical
simulations.  That simulations are consistent with numerous observed
galaxies increases our confidence in using bars to explain boxy/PS
bulges and figure-of-eight PVDs.

\section{Summary}
We have presented new spectroscopic and photometric observations of
the near edge-on spiral galaxy NGC~2683.  The long-slit PVD displays a
complex, figure-of-eight distribution of velocities, and the SparsePak
IFU velocity field shows characteristics of an oval distortion.
Isophotes in optical and near-IR images of the galaxy are boxy.  We
find an asymmetric enhancement in the light profile in slits placed
parallel to, and offset from, the major axis.

We argue that the kinematic and photometric signatures are evidence
that the galaxy hosts a bar.  We find that the data support a bar that
is viewed closer to side-on than end-on (i.e. closer to the galaxy
major axis than the minor axis). Based on kinematic and photometric
constraints, we determine that the bar is 6$\degr$ away from the major
axis on the sky.  Our results are consistent with previous
observations of NGC~2683 in the literature, as well as recent results
for M31, and add to the growing body of evidence linking B/PS bulges
and complex PVDs with bars in edge-on galaxies.

\acknowledgements 
We would like to thank the referee for helpful comments.  R.K.D. is
supported by an NSF Astronomy \& Astrophysics Postdoctoral Fellowship
under award AST 07-02496.  S.S.M. and M.J.Z. are supported by NSF
grant AST 05-05956.  We would like to thank L.~Athanassoula for
helpful conversations.  R.K.D. would also like to thank Brian
Marsteller and Misty Bentz for useful discussions.  This publication
makes use of data products from the Two Micron All Sky Survey, which
is a joint project of the University of Massachusetts and the Infrared
Processing and Analysis Center/California Institute of Technology,
funded by the National Aeronautics and Space Administration and the
National Science Foundation.

\end{document}